\documentclass[aps, pre, twocolumn, a4paper, superscriptaddress]{revtex4-2}

\usepackage{graphicx}
\usepackage{color}
\usepackage{amsmath, amssymb}
\usepackage{lmodern}
\usepackage[T1]{fontenc}
\usepackage[utf8]{inputenc}
\usepackage[english]{babel}
\usepackage{array}
\usepackage{multirow}
\usepackage[table]{xcolor}

\def\lsim{\mathrel{\rlap{\lower4pt\hbox{\hskip1pt$\sim$}}
    \raise1pt\hbox{$<$}}}       
\def\gsim{\mathrel{\rlap{\lower4pt\hbox{\hskip1pt$\sim$}}
    \raise1pt\hbox{$>$}}}       

\begin{document}

\title{Impact of parity in rock-paper-scissors type models}

\author{P.P. Avelino}
\affiliation{Instituto de Astrof\'{\i}sica e Ci\^encias do Espa{\c c}o, Universidade do Porto, CAUP, Rua das Estrelas, PT4150-762 Porto, Portugal}
\affiliation{Departamento de F\'{\i}sica e Astronomia, Faculdade de Ci\^encias, Universidade do Porto, Rua do Campo Alegre 687, PT4169-007 Porto, Portugal}
\author{B.F. de Oliveira}
\affiliation{Departamento de F\'{\i}sica, Universidade Estadual de Maring\'a, Av. Colombo 5790, 87020-900 Maring\'a, PR, Brazil}
\author{R.S. Trintin}
\affiliation{Departamento de F\'{\i}sica, Universidade Estadual de Maring\'a, Av. Colombo 5790, 87020-900 Maring\'a, PR, Brazil}

\begin{abstract}

We investigate the impact of parity on the abundance of weak species in the context of the simplest generalization of the rock-paper-scissors model to an arbitrary number of species --- we consider models with a total number of species ($N_S$) between 3 and 12, having one or more (weak) species characterized by a reduced predation probability (by a factor of ${\mathcal P}_w$ with respect to the other species). We show, using lattice based spatial stochastic simulations with random initial conditions, large enough for coexistence to prevail, that parity effects are significant. We find that the performance of weak species is dependent on whether the total number of species is even or odd, especially for $N_S \le 8$, with odd numbers of species being on average more favourable to weak species than even ones. We further show that, despite the significant dispersion observed among individual models, a weak species has on average a higher abundance than a strong one if ${\mathcal P}_w$ is sufficiently smaller than unity --- the notable exception being the four species case. 

\end{abstract}

\maketitle

\section{Introduction \label{sec1}}

Non-hierarchical predator-prey models are an important tool for  the understanding of the dynamics of complex biological systems involving a large number of species. Several models of this type have been proposed and investigated in the literature \cite{1996-Kobayashi-JPN-66-38, 1999-Tainaka-JTB-197-1, 2002-Sato-AMC-126-255, 2007-Szabo-PRE-76-051921, 2008-Szabo-PRE-77-011906, 2008-Peltomaki-PRE-78-031906,  2008-Szabo-PRE-77-041919, 2010-Nakagiri-ECO-5-241, 2011-Allesina-PNAS-108-5638, 2012-Avelino-PRE-86-031119, 2012-Li-PA-391-125, 2012-Roman-JSMTE-2012-p07014, 2012-Avelino-PRE-86-036112, 2013-Lutz-JTB-317-286, 2013-Roman-PRE-87-032148, 2014-Cheng-SR-4-7486, 2014-Szolnoki-JRSI-11-0735, 2016-Kang-Entropy-18-284, 2016-Roman-JTB-403-10, 2017-Brown-PRE-96-012147, 2017-Park-SR-7-7465, 2017-Bazeia-EPL-119-58003, 2017-Souza-Filho-PRE-95-062411, 2018-Shadisadt-PRE-98-062105, 2019-Avelino-PRE-99-052310, 2020-Szolnoki-EPL-131-68001}, most of them considering three basic interactions --- predator-prey,  reproduction and mobility --- usually assumed to happen with the same probability for all species (reproduction and mobility) or pairs of non-minimally interacting species (predator-prey). Parity --- the even or odd nature of the total number of species --- has been observed to play an important role in many of these models. It can affect not only the dynamics of the network, and consequently coexistence, but also the properties of the geometrical patterns associated to the resulting dynamical structures or the symmetric/asymmetric evolution of the interface profiles separating different domains \cite{1996-Kobayashi-JPN-66-38, 1999-Tainaka-JTB-197-1, 2002-Sato-AMC-126-255, 2007-Szabo-PRE-76-051921, 2008-Szabo-PRE-77-011906, 2008-Peltomaki-PRE-78-031906, 2010-Nakagiri-ECO-5-241,  2012-Avelino-PRE-86-036112}.

In \cite{2012-Avelino-PRE-86-036112} a generalization of the  rock-paper-scissors (RPS) model to an arbitrary number of species has been developed. This family of models generates dynamical spiral structures with a number of arms equal to the number of species. If the predation, reproduction and mobility rates do not vary from species to species, the resulting average density is the same for all species, assuming that the simulation box is large enough for coexistence to prevail --- for smaller simulation boxes, the survival probabilities are also equal for all species, 
assuming  unbiased initial conditions. 

In realistic biological systems the species strength is not expected to be the same for all species, which can affect both the population sizes and the chances of survival of the different species. Species with a reduced predation probability are usually referred to as weak. Nevertheless, it has been shown, in the context of three species RPS models, that weak species may often have a strong performance both in terms of population abundance and survival probability \cite{2001-Frean-PRSLB-268-1323,2009-Berr-PRL-102-048102, 2019-Avelino-PRE-100-042209, arxiv} (see also \cite{2019-Menezes-EPL-126-18003}). Recent research \cite{2020-Liao-N-11-6055} on the dynamics of three strains of \textit{E. coli} interacting cyclically also concluded for the dominance of the weakest strain. Still, no systematic difference in the global performance of weak and strong species has been found in the context of RPS models with four species \cite{2020-Avelino-PRE-101-062312}. This result raises the following question: is parity a key factor on the performance of weak and strong species?

In this paper we consider the simplest generalization of the spatial stochastic RPS model to an arbitrary number of species proposed in \cite{2012-Avelino-PRE-86-036112}, but shall relax the assumption that all species have equal strength. We investigate the impact that a reduction of the predation probability of some of the species --- the weak ones --- has on the overall dynamics of the network and, in particular, on the abundance of such species. We shall consider models with a total number of species $N_S$ between 3 and 12, paying particular attention to the impact of parity on the overall abundance of weak and strong species. 

\begin{figure}[t]
    \centering
    \includegraphics[width=8.2cm]{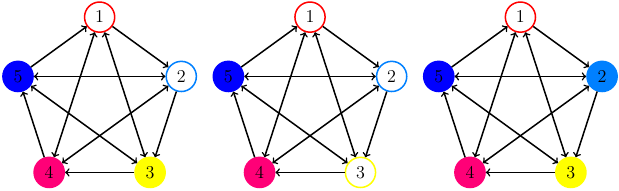}
    \includegraphics[width=8.2cm]{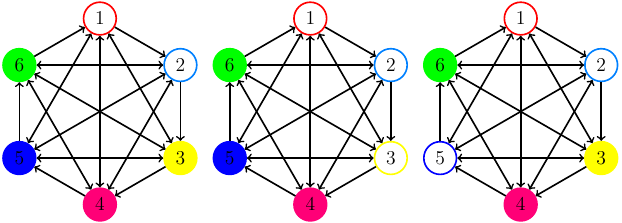}
    \includegraphics[width=8.2cm]{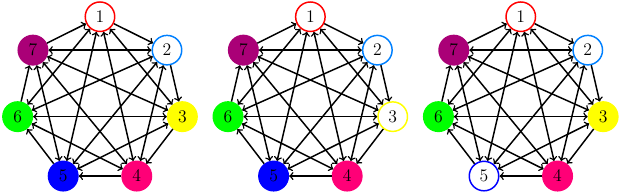}
    \includegraphics[width=8.2cm]{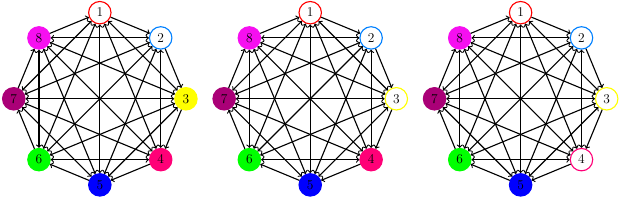}
    \caption{Predator-prey interactions of a sample of models with $5$, $6$, $7$ and $8$ species studied in the present paper (corresponding the first three models listed in Table I with that number of species). Filled and open circles represent weak and strong species, respectively.}
    \label{fig1}
\end{figure}

\section{RPS type models with $N_S$ species \label{sec2}}

Here, we shall briefly describe the simplest generalization of the spatial stochastic RPS model to $N_S$ species (May-Leonard formulation) \cite{2012-Avelino-PRE-86-036112}. In this model the different species are labelled by $i$ (or $j$) with $i,j = 1,...,N_S$, and modular arithmetic, where integers wrap around upon reaching $1$ or $N_S$, is used (the integers $i$ and $j$ represent the same species whenever $i = j \ {\rm mod} \ N_S$, where mod denotes the modulo operation). We shall perform spatial stochastic simulations on a square lattice with $N^2$ sites and periodic boundary conditions, employing a  May-Leonard formulation in which every site is either occupied by a single individual of one of the $N_S$ species or empty (an empty site is represented by a `$0$'). The number of individuals of the species $i$ and the number of empty sites is denoted by $I_i$ and $I_0$, respectively --- the density of individuals of the species $i$ and the density of empty sites are defined respectively by $\rho_i=I_i/N^2$ and $\rho_0 = I_0/N^2$. The allowed  interactions are predation [$i\ (i+1) \to i\ 0$], reproduction [$i\ 0 \to i\ i$], and mobility [$i\ \odot \to \odot\ i$], where $\odot$ represents either an individual of any species or an empty site. Reproduction and mobility interactions occur, respectively, with probabilities $r$ and $m$ (assumed to be the same for all the species). In our baseline model the predation probability $p$ is the same for all possible predator-prey interactions. However, in this paper we shall investigate the dynamical impact of a reduction of the predation probability the weak species by a factor of $\mathcal{P}_w \in \ ]0,1[$ (the other species are sometimes referred to as strong). 

The predator-prey interactions of a sample of models with $5$, $6$, $7$ and $8$ species studied in the present paper (corresponding to the first three models listed in Table I with that number of species) are represented in Fig. \ref{fig1}, with filled and open circles representing weak and strong species, respectively. The one-sided arrows represent one-directional predator-prey interactions between species $i$ and $i+1$, while the double sided arrows represent the possible bi-directional predator-prey interactions between species $i$ and $i+2$, $\dots$, $i+N_S-2$. The list of all the different models with $N_S=5,6,7,8$ is displayed in Table I, where the digits identifying each model represent the weak species. For each value of $N_S$, the models listed in Table I have been ordered in decreasing order (from left to right, and then top to bottom) of the value of ${\mathcal A}_w$ for $\mathcal{P}_w=0.5$ shown in  Fig.~\ref{fig6} (the relevance of this ordering will become clear later on, when discussing the results shown in Fig.~\ref{fig6}). Notice that there are in general several combinations of digits corresponding to a single model. In particular, any permutation among weak or among strong species results in an equivalent model. Also, relabeling all the species $i$ as $i+n$, where $n$ is an integer, does not lead to a different model. Hence, only one equivalent combination is listed in Table I. For example, models $21$, $23$, $32$, $34$, $43$, $45$, and $54$ are all equivalent to $12$ for any $N_S \ge 5$, which is the digit combination which appears in table I (also notice that, for $N_S=5$, models $15$ and $51$ would also both be equivalent to $12$).

\begin{table}[t]
\begin{tabular}{|m{1.08cm}|m{1.25cm}|m{1.25cm}|m{1.25cm}|m{1.25cm}|m{1.25cm}|}
	\hline 
	\multirow{2}{*}{$N_S = 5$} 
	& 12      & 123     & 1       & 1234    & 134     \\
	\cline{2-6}
	& 13      & \multicolumn{3}{r}{} &  \\
	\hline \hline
	\multirow{3}{*}{$N_S = 6$} 
	& 12      & 123     & 125     & 14      & 1234    \\
	\cline{2-6}
	& 1245    & 1       & 12345   & 124     & 12355   \\
	\cline{2-6}
	& 13       & 135   & \multicolumn{2}{r}{} &  \\
	\hline \hline
	\multirow{4}{*}{$N_S = 7$} 
	& 12      & 123     & 125     & 1245    & 1236    \\
	\cline{2-6}
	& 126     & 1234    & 12356   & 1       & 12345   \\
	\cline{2-6}
	& 124     & 14      & 1235    & 123456  & 12346   \\
	\cline{2-6}
	& 1246    & 13      & 135    & \multicolumn{1}{r}{} &  \\
	\hline \hline
	\multirow{7}{*}{$N_S = 8$} 
	& 12      & 123     & 1234    & 127     & 1256    \\
	\cline{2-6}
	& 1245    & 125     & 1236    & 12356   & 12456   \\
	\cline{2-6}
	& 12347   & 1237    & 126     & 1257    & 123467  \\
	\cline{2-6}
	& 1       & 14      & 124     & 12457   & 12345   \\
	\cline{2-6}
	& 1247    & 123456  & 123567  & 1235    & 12346   \\
	\cline{2-6}
	& 1234567 & 136     & 123457  & 12357   & 1246    \\
	\cline{2-6}
	& 15      & 13      & 135     & 1357    &         \\
	\hline 
\end{tabular}
\caption{List of all the different models with $N_S=5,6,7,8$. The digits identifying each model represent the weak species. The models have been ordered in decreasing order (from left to right, and then top to bottom) of the value of ${\mathcal A}_w(\mathcal{P}_w=0.5)$ shown in the four panels .}
\end{table}

At every simulation step, the algorithm randomly picks an occupied site to be the active one, randomly selects one of its adjacent neighbour sites to be the passive one, and randomly chooses an interaction to be executed by the individual at the active position: predation, mobility or reproduction with probabilities $p$, $m$ and $r$, respectively --- in this paper we use the von Neumann neighbourhood (or 4-neighbourhood) composed of a central cell (the active one) and its four non-diagonal adjacent cells (it has been shown in \cite{2019-Avelino-PRE-100-042209}, in the context of a three species model, that a Moore neighbourhood leads to the same qualitative results). These three actions are repeated until a possible interaction is selected --- note that the interaction cannot be carried out whenever predation is selected and the passive is not a prey of the active, or if reproduction is selected and the passive is not an empty site. A generation time (our time unit) is defined as the time necessary for $N^2$ successive interactions to be completed. 

\begin{figure}[t]
	\centering
	\includegraphics{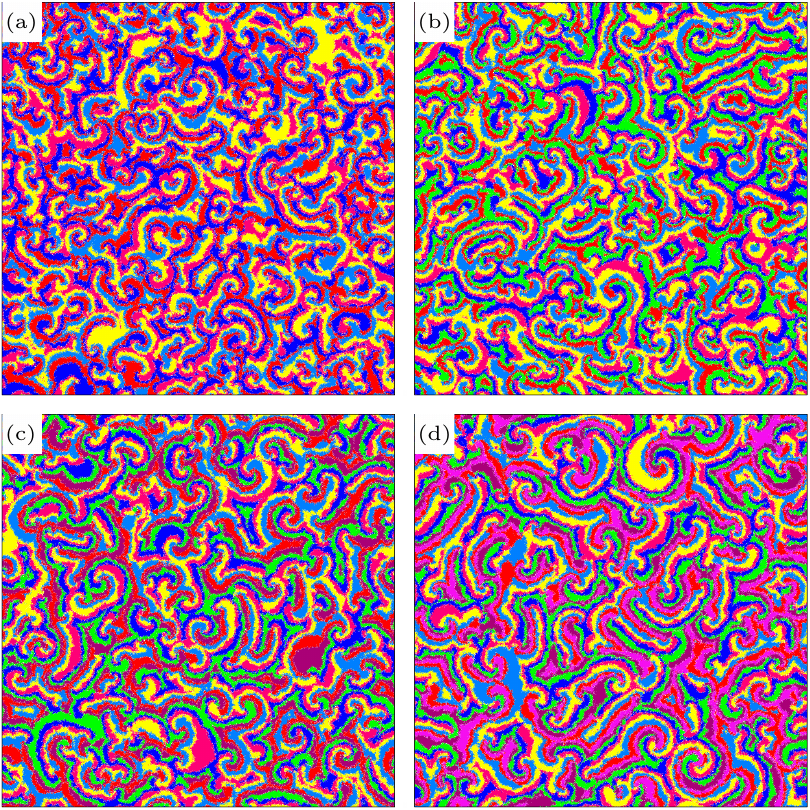}
	\caption{The panels display snapshots of the spatial distribution of the different species on a $1000^2$ lattice at $t=1.1 \times 10^4$ for realizations of the spatial stochastic RPS type models with: (a) $N_S=5$, (b) $N_S=6$, (c) $N_S=7$ and (d) $N_S=8$ (the other model parameters are $m=0.2$, $p=0.4$, $r=0.4$, and $\mathcal{P}_w=1$). Notice the appearance of spiral patterns with a number of arms equal to the number of species $N_S$.}
	\label{fig2}
\end{figure}

\section{Results \label{sec3}}

In this section we present the results of $1000^2$ spatial stochastic numerical simulations, considering models with a number of species in the interval $[3,12]$ and different values of $\mathcal{P}_w$. The parameters $m=0.2$, $p=0.4$, $r=0.4$ are assumed in all simulations. 

Fig.~\ref{fig2} displays the distribution of the different species on a square lattice after $1.1 \times 10^4$ generations for a model in which all the species have the same strength ($\mathcal{P}_w=1.0$). The number of species are (a) $N_S=5$ (top left panel), (b) $N_S=6$ (top right panel), (c) $N_S=7$ (bottom left panel) and (d) $N_S=8$ (bottom right panel). Spiral patterns with a number of arms equal to the number of species are present in all the simulations. Also, no clear predominance of one species over the others is observed in any of the snapshots (as expected, since $\mathcal{P}_w=1.0$).

\begin{figure}[t]
	\centering
	\includegraphics{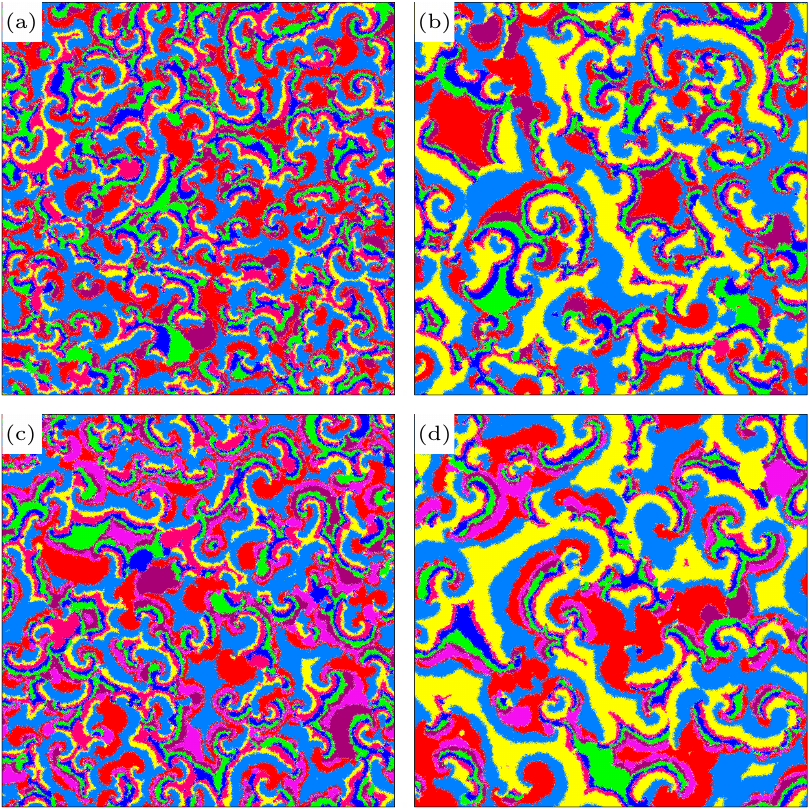}
	\caption{The same as in Fig.~\ref{fig2}c and Fig.~\ref{fig2}d (RPS type models with seven and eight species), but now considering $\ \mathcal{P}_w=0.5$: $(a)$ $N_S=7$, model $1$; $(b)$ $N_S=7$, model $12$; $(c)$ $N_S=8$, model $1$; $(d)$ $N_S=8$, model $12$. Notice the significant differences between the shapes and characteristic sizes of the domains associated to the different species.}
	\label{fig3}
\end{figure}

Fig.~\ref{fig3} is analogous to the bottom panels of Fig.~\ref{fig2} (Figs.~\ref{fig2}c and ~\ref{fig2}d, which contemplate cases with seven and eight species, respectively), but now considering models $1$ and $12$, and $\mathcal{P}_w=0.5$: $(a)$ $N_S=7$, model $1$; $(b)$ $N_S=7$, model $12$; $(c)$ $N_S=8$, model $1$; $(d)$ $N_S=8$, model $12$. Contrarily to Fig.~\ref{fig2}, Fig.~\ref{fig3} shows that some species are more dominant than others. Also, although the spiral patterns are still easily recognizable, significant differences exist between the shapes and characteristic sizes of the domains associated to the different species. This is true both for the models with one weak species (left panels) and two weak species (right panels), but more so in the latter. 

\begin{figure}[t]
	\centering
	\includegraphics{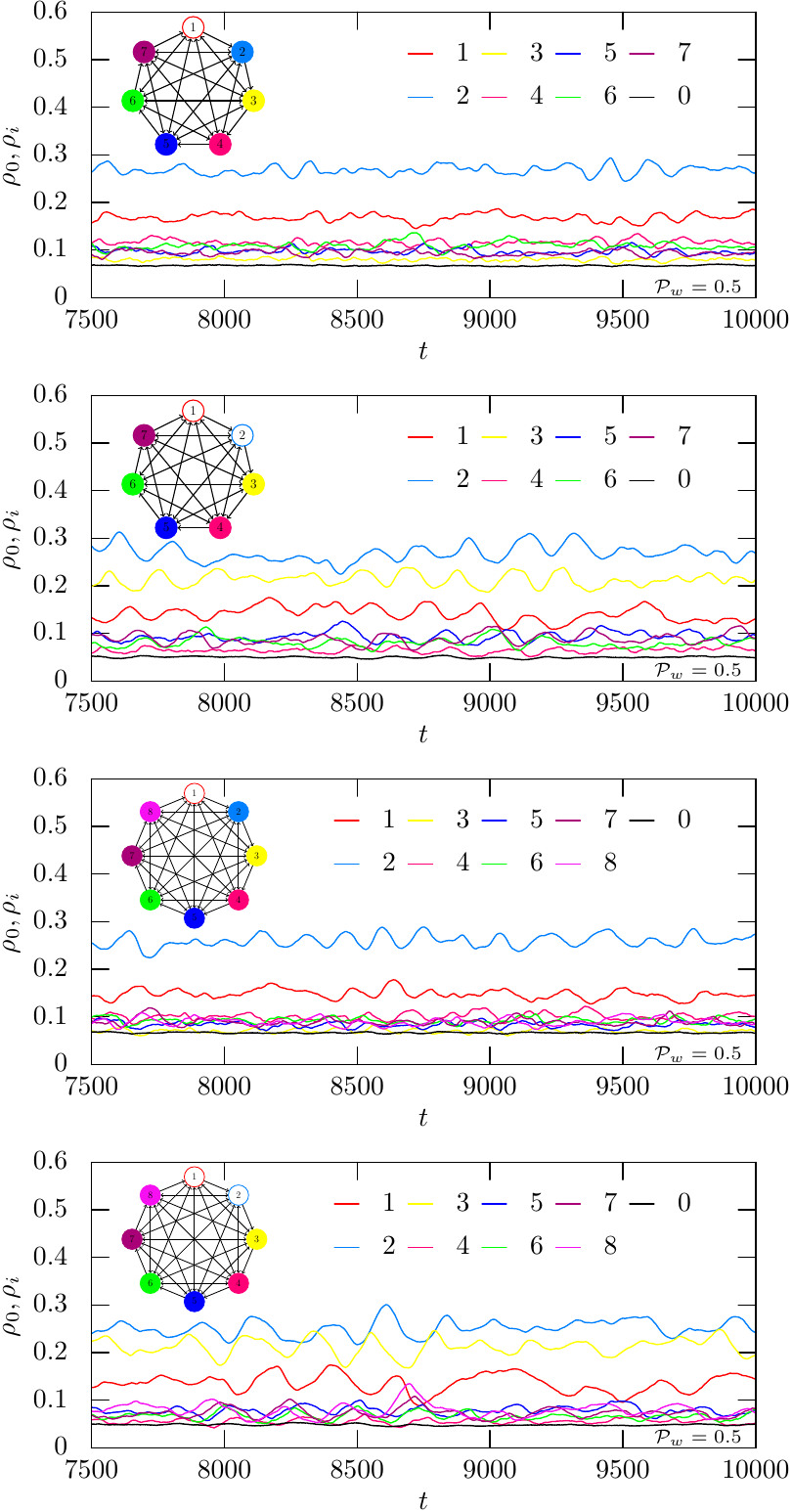}
	\caption {The evolution of the density of the different species and empty sites ($\rho_i$ and $\rho_0$) over time for realizations of spatial stochastic RPS type models with $7$ and $8$ species considered in Fig.~\ref{fig3}a, Fig.~\ref{fig3}b, Fig.~\ref{fig3}c and Fig.~\ref{fig3}d (from top to bottom, respectively). Notice the impact that the inclusion of one or two weak species has on the abundance of the different species.}
	\label{fig4}
\end{figure}

Fig.~\ref{fig4} displays the evolution of the densities of the different species and empty spaces for the realizations of spatial stochastic RPS type models with $7$ and $8$ species considered in ~\ref{fig3}: $(a)$ $N_S=7$, model $1$ (1st panel); $(b)$ $N_S=7$, model $12$ (2nd panel); $(c)$ $N_S=8$, model $1$ (3rd panel); $(d)$ $N_S=8$, model $12$ (4th panel). The impact that the reduced predation probability  of one or two weak species has on the abundance of the different species is qualitatively similar for models with seven and eight species. In model $1$, with a single weak species, the most abundant species is the prey of the weak species (represented by a blue filled circle), followed by the weak species (represented by a red open circle). In model $12$, with two weak species, the most abundant species is one of the weak species (the one represented by a blue open circle), followed by its prey (represented by a yellow filled circle), and then by the other weak species (represented by a red open circle).

In order to quantify the impact of the species strength on their overall abundance, we define the average density of weak and strong species as 
\begin{equation}
\langle \rho_w \rangle = \frac{1}{\#W} \sum_{i \in W} \langle \rho_i \rangle\,, \quad \langle \rho_s \rangle = \frac{1}{\#S} \sum_{i \in W} \langle \rho_i \rangle\,,
\end{equation}
where $W$ and $S$ are, respectively, the sets whose elements are the weak and the strong species, and $\#$ is used to represent the number of elements of each set. Let us also define the parameter 
\begin{equation}
{\mathcal A}_w=\frac{\langle \rho_w \rangle-\langle \rho_s \rangle}{\max(|\langle \rho_w \rangle|,|\langle \rho_s \rangle|)}\,,
\end{equation}
whose absolute value represents the relative advantage (if ${\mathcal A}_w>0$) or disadvantage (if ${\mathcal A}_w<0$) in being a weak species.

\begin{figure}[t]
	\centering
	\includegraphics[width=8.5cm]{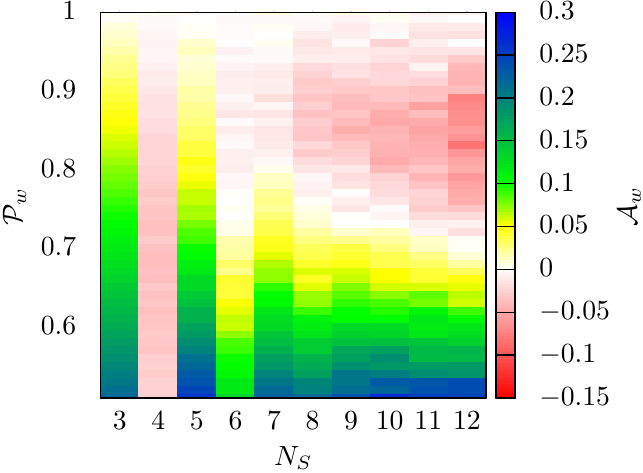}
	\includegraphics[width=8.5cm]{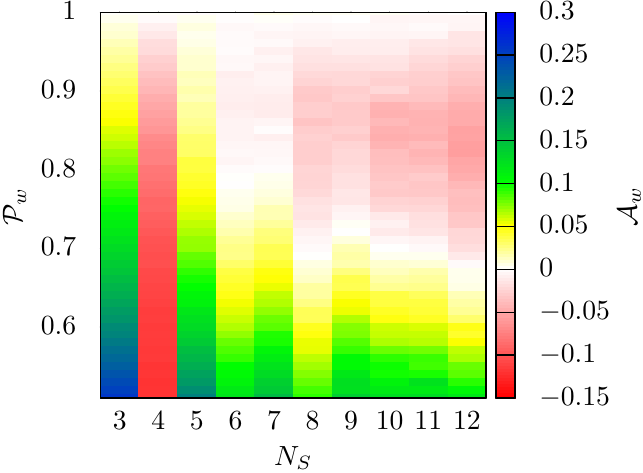}
	\caption{The relative advantage in being a weak species ${\mathcal A}_w$ (or disadvantage if $\mathcal{A}_w < 0$) as a function of $\mathcal{P}_w$ and $N_S$ for models with one (top panel) or two (bottom panel) weak species, considering a total number of species $N_S$ between 3 and 12 and $\mathcal P_w$ between $0.5$ and $1$ (in the bottom panel the average value of $\mathcal{A}_w$ among models with two weak species is considered). Notice the impact of parity, specially for $N_{S} \le 8$.}
	\label{fig5}
\end{figure}

\begin{figure*}[t]
	\centering
	\includegraphics[width=17cm]{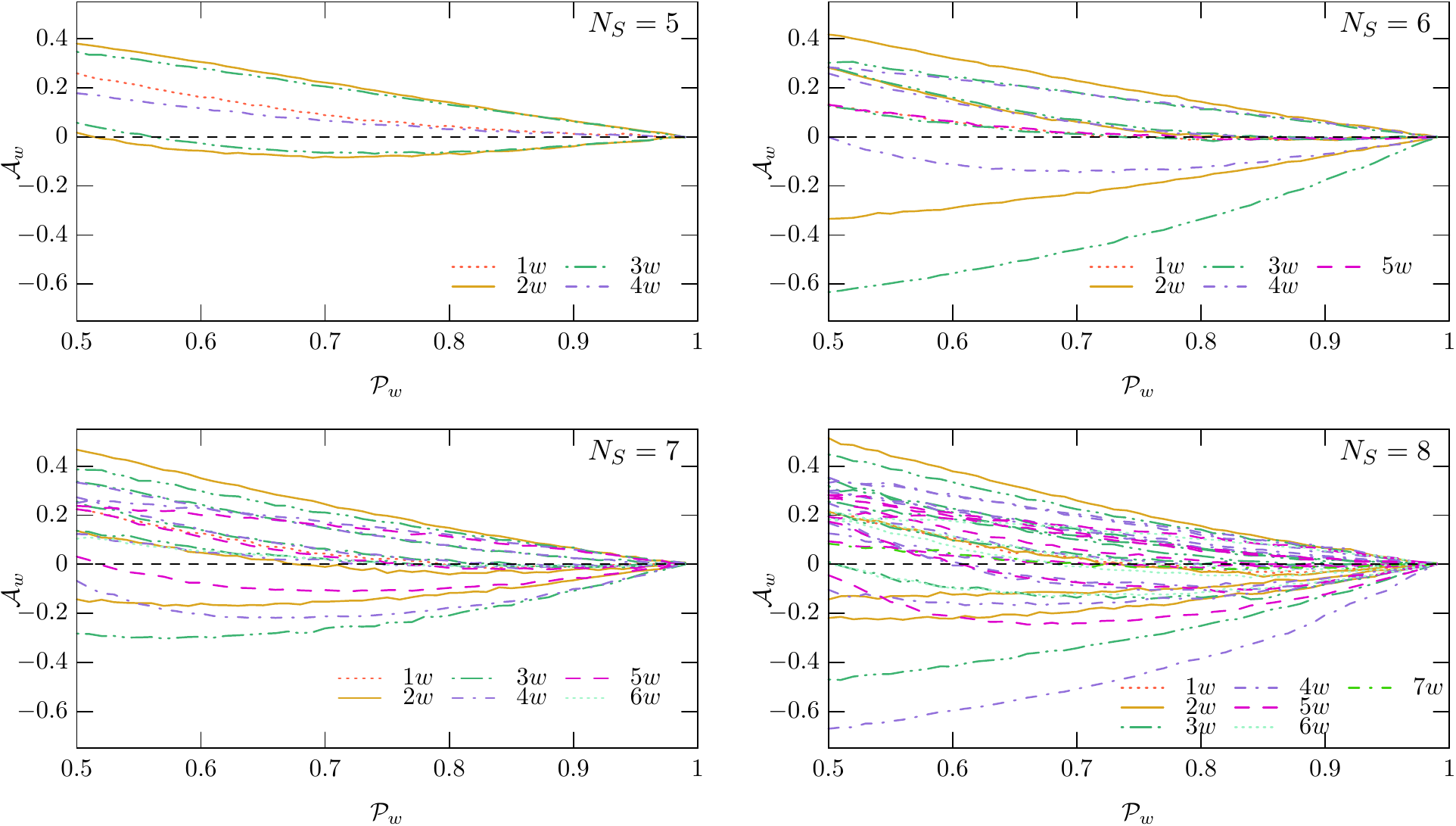}
	\caption{The relative advantage in being a weak species ${\mathcal A}_w$ (or disadvantage if $A_w < 0$) as a function of $\mathcal{P}_w$ for the cases with $N_{S}=5$, $N_{S}=6$, $N_{S}=7$ and $N_{S}=8$. The dashed black line represents $\mathcal{A}_w=0$. The different colors and line types represent models with a different number of weak species.}
	\label{fig6}
\end{figure*}

\begin{figure}[t]
	\centering
	\includegraphics[width=8.5cm]{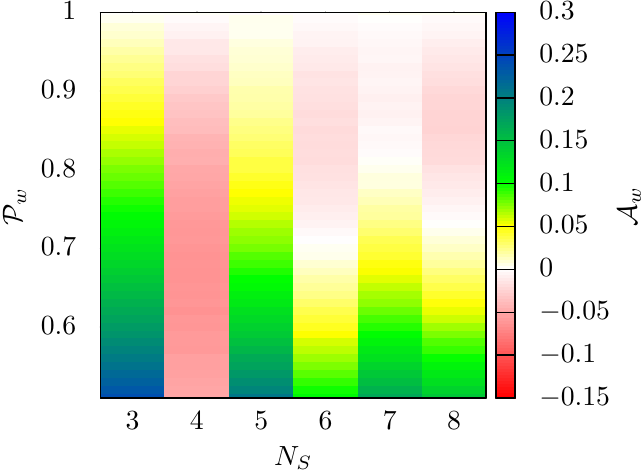}
	\caption{The average relative advantage in being a weak species ${\mathcal A}_w$ as a function of $\mathcal{P}_w$ and $N_S$, for models with a total number of species between 3 and 8. Again, the impact of parity is significant for $N_{S} \le 8$, with odd numbers of species being more favourable to weak species than even ones.}
	\label{fig7}
\end{figure}

The top panel of Fig.~\ref{fig5} shows the relative advantage in being a weak species ${\mathcal A}_w$ (or disadvantage if $\mathcal{A}_w < 0$) as a function of $\mathcal{P}_w$ and $N_S$, for models with one weak species, considering a total number of species $N_S$ between 3 and 12 and $\mathcal P_w$ between $0.5$ and $1$. The bottom panel displays the average relative advantage among all two weak species models with $N_S$ species (also represented by $\mathcal{A}_w$ for simplicity of notation) as a function of $\mathcal{P}_w$ and $N_S$. In both models the impact of parity is significant at least up to $N_S=8$, with odd numbers of species being more favourable to weak species than even ones. In these models the weak species generally have a significant an advantage if ${\mathcal P}_w$ is sufficiently smaller than unity, the exception being the models with $N_S=4$ (this particular case has been investigated in detail in \cite{2020-Avelino-PRE-101-062312}).

Taking into account that parity effects are most noticeable for $N_S \le 8$, we shall now consider the models with a number of species $N_S$ between $5$ and $8$ one by one (the models with $3$ and $4$ species have already been investigated in detail in \cite{2019-Avelino-PRE-100-042209, arxiv} and \cite{2020-Avelino-PRE-101-062312}, respectively). Fig.~\ref{fig6} displays the relative advantage in being a weak species ${\mathcal A}_w$ (or disadvantage if $A_w < 0$) as a function of $\mathcal{P}_w$ for all models with $N_{S}=5$, $N_{S}=6$, $N_{S}=7$ and $N_{S}=8$ (the dashed black line represents $\mathcal{A}_w=0$). The different colors and line types represent models with a different number of weak species. In order to allow for a better identification of the models in Fig.~\ref{fig6}, the models in Table I have been ordered in decreasing order (from left to right, and then top to bottom) of the value of ${\mathcal A}_w$ obtained for $\mathcal{P}_w=0.5$.

Fig.~\ref{fig6} shows that for any particular $N_S$ between $5$ and $8$ there is a significant dispersion of the curves of ${\mathcal A}_w({\mathcal P}_w)$. This is true even if only models with a fixed number of weak species are selected. Fig.~\ref{fig6} also shows that, for any value of $N_S$ between $5$ and $8$, the average abundance of a weak species is higher than that of a strong one in most models. Still, there are a few models with an even number of species which are highly adverse for weak species (e.g. models $135$ and $1357$, respectively for $N_S=6$ and $N_S=8$) --- this does not happen if the number of species is odd. 

Fig.~\ref{fig7} shows the average relative advantage as a function of the number of species and a total number of species $N_S$ between 3 and 8. Again the impact of parity is noticeable, with odd numbers of species being on average more favourable to weak species than even ones if ${\mathcal P}_w$ is sufficiently smaller than unity. The results displayed in Fig.~\ref{fig7} for $N_S \le 8$ are qualitatively similar to those of Fig.~\ref{fig5}.

\section{Conclusions \label{conc}}

In this paper we investigated the dynamics of RPS type models with a total number of species between 3 and 12, in the presence of one or more weak species. We showed that parity effects are significant, with the abundance of weak species having a significant dependence on whether the number of species is even or odd for $N_S \le 8$. 

We have shown that, unlike in the case of RPS models with three different species \cite{2019-Avelino-PRE-100-042209, arxiv}, the relative advantage in being a weak species may vary significantly from model to model in models with more than four species and a fixed value of the reduced predation probability --- a significant dispersion is observed even among models with a fixed number of weak species. These results are in agreement with the findings of \cite{2020-Avelino-PRE-101-062312} where only models with four species were considered. 

Notwithstanding the large dispersion among models, we found that a weak species has on average a significant advantage over a strong one if ${\mathcal P}_w$ is sufficiently smaller than unity, the only exception being the four species case. Still, we have shown that parity plays a key role, with odd numbers of species being on average more favourable to weak species than even ones in terms of the overall abundance.

\begin{acknowledgments}
P.P.A. acknowledges the support Fundação para a Ciência e a Tecnologia (FCT) through the research grants UIDB/04434/2020, UIDP/04434/2020. B.F.O. and R.S.T. thank CAPES - Finance Code 001, Funda\c c\~ao Arauc\'aria, and INCT-FCx (CNPq/FAPESP) for financial and computational support.
\end{acknowledgments}

%
\end{document}